\documentclass[11pt,preprintnumbers,aps,amssymb,nofootinbib,amsmath,notitlepage,tightenlines,prd,superscriptaddress]
{revtex4-1}
\usepackage{epsfig,epsf}
\usepackage{bm} 
\usepackage{color}
\definecolor{refkey}{gray}{.50} 
\definecolor{labelkey}{gray}{.05} 
\usepackage{slashed}
\usepackage{xcolor} 
\usepackage{mathtools}
\usepackage{mathrsfs}
\usepackage{graphicx}
\usepackage[caption=false]{subfig}
\usepackage{relsize}	
\usepackage{soul}
\usepackage{hyperref}
\hypersetup{colorlinks=true,linkcolor=darkgreen,anchorcolor=blue,citecolor=darkgreen, filecolor=blue,urlcolor=blue,bookmarksnumbered=true,
pdfview=FitB
} 

\colorlet{darkgreen}{green!60!black}
\colorlet{brightyellow}{yellow!75!red}
\colorlet{orange}{red!50!yellow}
\colorlet{darkblue}{blue!80!green}
\colorlet{darkred}{red!80!black}
\colorlet{greenblue}{green!50!blue}
\colorlet{darkgray}{gray!40!black}

\makeatletter

\newcommand{\Rmnum}[1]{\expandafter\@slowromancap\romannumeral #1@}
\makeatother

\newcommand{\beq}{\begin{equation}}
\newcommand{\beql}[1]{\begin{equation}\label{#1}}
\newcommand{\eeq}{\end{equation}}
\def\bal#1\gal{\begin{align}#1\end{align}}
\newcommand{\ball}[1]{\bal\label{#1}}
\newcommand{\eq}[1]{(\ref{#1})}
\newcommand{\fig}[1]{Fig.~\ref{#1}}
\renewcommand{\sec}[1]{Sec.~\ref{#1}}
\newcounter{topiccounter}
\renewcommand{\b}[1]{{\bm #1}} 

\begin{document}

\title{Electrodynamics of dual superconducting chiral medium}

\author{Yang Li}

\affiliation{Department of Physics and Astronomy, Iowa State University, Ames, IA 50011, USA}
\affiliation{Department of Physics, College of William \& Mary, Williamsburg, VA 23187, USA}

\author{Kirill Tuchin}

\affiliation{Department of Physics and Astronomy, Iowa State University, Ames, IA 50011, USA}

\date{\today}


\begin{abstract}

We study the electrodynamics of a chiral medium with electric and magnetic charges using the effective Maxwell-Chern-Simons theory extended to include the magnetic current. The exchange of helicity between the chiral medium and the magnetic field, known as the inverse cascade, is controlled by the chiral anomaly equation. In the presence of the magnetic current, the magnetic helicity is dissipated, so that the inverse cascade stops when the magnetic helicity vanishes while the chiral conductivity reaches a non-vanishing stationary value satisfying $\sigma_\chi^2< 4\sigma_e\sigma_m$, where $\sigma_e$, $\sigma_m$ and $\sigma_\chi$ are electric, magnetic and chiral conductivities respectively.  We argue that this state is superconducting and exhibits the Meissner effect for both electric and magnetic fields. Moreover, this state is stable with respect to small magnetic helicity fluctuations; the magnetic helicity becomes unstable  only when the inequality mentioned above is violated. 

\end{abstract}

\maketitle

\section{Introduction}\label{sec:i}

Classical electromagnetic field in a medium with chiral anomaly is described by a system of  Maxwell equations and the chiral anomaly equation \cite{Adler:1969gk,Bell:1969ts} known as the 
Maxwell-Chern-Simons (MCS) theory \cite{Wilczek:1987mv,Carroll:1989vb, Sikivie:1984yz,Kharzeev:2009fn}. The chiral anomaly equation controls exchange of helicity between the field and medium such that the total helicity is conserved. The resulting non-trivial evolution of the magnetic field topology has been a subject of recent interest \cite{Joyce:1997uy,Boyarsky:2011uy,Hirono:2015rla,Hirono:2016jps,Tashiro:2012mf,Manuel:2015zpa,Kedia:2013bw,Hoyos:2015bxa,Rogachevskii:2017uyc,Pavlovic:2016gac,Xia:2016any,Yamamoto:2016xtu}  motivated by phenomenological applications in nuclear physics, condensed matter physics and cosmology \cite{Kharzeev:2013ffa}.

A distinctive feature of the MCS theory is the emergence of the soft magnetic field modes exponentially growing in time \cite{Tuchin:2014iua,Manuel:2015zpa,Joyce:1997uy,Boyarsky:2011uy,Kharzeev:2013ffa,Khaidukov:2013sja,Kirilin:2013fqa,Avdoshkin:2014gpa,Akamatsu:2013pjd,Dvornikov:2014uza,Buividovich:2015jfa,Sigl:2015xva,Xia:2016any,Kirilin:2017tdh}. These unstable modes 
transfer helicity from the medium to the field in a process known as the inverse cascade \cite{Biskamp,Boyarsky:2011uy}. Eventually, however, the helicity conservation puts a cap on the inverse cascade \cite{Kaplan:2016drz,Tuchin:2017vwb}.

It has been argued in \cite{Liao:2006ry,Lopez-Ruiz:2016bjl,Ratti:2008jz,Lublinsky:2009iu,Xu:2014tda} that magnetic monopoles play an important role in quark-gluon plasma dynamics. Magnetic monopoles also often appear in cosmological models \cite{Preskill:1984gd} and even in condensed matter physics \cite{Castelnovo:2007qi}. This motivates us to consider the MCS theory with dynamical magnetic monopoles (MCSm). That the magnetic monopoles are expected to have non-trivial effects on the magnetic field can be seen from the fact that the dual transformation generates in the Lagrangian the same $CP$-odd term as the chiral anomaly. In particular, the magnetic current, while being energy non-dissipative,  causes dissipation of the total helicity. The main goal of this paper is to uncover the main features of the chiral magnetic dynamics with magnetic monopoles. 

The paper is organized as follows. In \sec{sec:a} we formulate the equations of the MCSm theory and  analyze their main properties. Our main assumption is the linear medium response that is characterized by the electric and magnetic conductivities $\sigma_e$ and $\sigma_m$. 
We observe the emergence of the superconducting phase when $\sigma_\chi^2<4\sigma_e\sigma_m$ and formulate the corresponding London equations \eq{m8},\eq{m9} in \sec{sec:meissner}. In \sec{sec:ca} we analyze the late-time dynamics of the MCSm system, in particular, its evolution towards a stationary state.  We argue that the magnetic helicity must exponentially decay due to the helicity dissipating magnetic current. The chiral conductivity $\sigma_\chi$ also decays owing to the inverse cascade as mentioned above. However, in the presence of the magnetic current, the inverse cascade may be terminated before the chiral conductivity turns zero. Therefore, the chiral conductivity approaches a finite stationary value $\sigma_\infty$ while the magnetic helicity is completely dissipated. In \sec{sec:d} we investigate the dispersion relation of the magnetic field modes and point out the conditions under which the magnetic field (and magnetic helicity) is unstable. In our context, the term ``instability" means that a small fluctuation of the field triggers its exponential growth, even though eventually it decays as a result of the magnetic helicity non-conservation.  We show that the stability condition coincides with the condition for the existence of the superconductivity. In order to develop a clearer understanding of the time evolution of the magnetic field and the chiral conductivity, we employ in \sec{sec:d+}  the Fastest Growing State (FGS) model \cite{Tuchin:2017vwb} which assumes that the magnetic helicity at later times is driven by a mode with the exponentially largest  growth rate. Using this model we perform in \sec{sec:e}  a detailed investigation of the time-evolution of the MCSm theory. We argue that after undergoing an inverse cascade the system settles to the superconducting phase. This is the main result of our paper. We conclude with a discussion in \sec{sec:s}.

\section{Maxwell-Chern-Simons theory with magnetic monopoles }\label{sec:a}

\subsection{Maxwell and the chiral anomaly equations}\label{sec:a0}

A plasma of electric and magnetic charges with chiral anomaly is governed by the following generalization of the Maxwell equations  \cite{Wilczek:1987mv,Carroll:1989vb, Sikivie:1984yz,Kharzeev:2009fn}:
\bal
&\b \nabla\cdot \b B=0\,, \label{a3}\\
& \b \nabla\cdot \b E=0\,,  \label{a4}\\
-& \b \nabla \times \b E= \partial_t \b B+\b j_m\,,\label{a5}\\
& \b \nabla \times \b B= \partial_t \b E+ \b j_e + \sigma_\chi \b B\,,\label{a6}
\gal
where $\b j_m$ is the magnetic current density and  $\sigma_\chi$ is assumed to depend only on time.  We neglected the electric and magnetic polarization of the plasma, which is a small effect for good conductors and consider the plasma to be electrically and magnetically neutral. 
Assuming  the linear response $\b j_e= \sigma_e\b E$, $\b j_m= \sigma_m\b B$ with constant electric and magnetic conductivities we can derive, using \eq{a3}--\eq{a6},  an equation for the magnetic field\footnote{ Magnetic field is supposed to be not very strong, so that the Larmor radius is much larger than the Debye radius $r_D$, which guarantees that the kinetic coefficients do not depend on $B$. For relativistic plasmas at temperature $T$ this amounts to  $eB \ll  r_D T$.   }
\ball{a9}
-\nabla^2 \b B+\partial_t^2 {\b B}=-(\sigma_e+\sigma_m)\partial_t{ \b B}- \sigma_e\sigma_m\b B+\sigma_\chi(t) \b \nabla\times \b B\,.
\gal
In view of \eq{a3} we can introduce the vector potential  $\b A$ as $\b B= \b\nabla\times \b A$. 
Since the Bianchi identity is violated in the presence of  the magnetic current, the relationship between the electric field and the vector potential is modified as compared to the Maxwell theory. 
One can check that 
\ball{a11}
\b E= -\partial_t\b A-\sigma_m\b A\,,
\gal
satisfies the modified Faraday's law \eq{a5} in the Coulomb gauge $\b \nabla\cdot \b A=0$. We note that the vector potential $\b A$ obeys the same equation \eq{a9} as the magnetic field.  

The relationship  \eq{a11} between the electric field and the vector potential is not unique. One can add on its right-hand-side a gradient of any scalar function $\phi$. The choice of $\phi$  is dictated by the requirement of  the gauge-invariance of \eq{a11}. Equations such as \eq{a11} appear in the theory of the superconductivity and indicate the necessity to introduce the magnetic monopole condensate. The condensate contributes to the right-hand-side of \eq{a11} a term proportional to the gradient of its phase $\phi$ which restores the gauge invariance. The term  $-\sigma_m\b A$ in \eq{a11} and the term proportional to $\b\nabla \phi$ make up  the supercurrent.  Not surprisingly, the supercurrent induces the Meissner effect  discussed in the next sub-section.   Throughout the paper we assume the gauge condition $\phi=0$ (the unitary gauge).

The time-evolution of the chiral conductivity is governed by the chiral anomaly equation. At high temperatures it can be written as \cite{Hirono:2015rla,Tuchin:2017vwb}
\ball{a13}
\partial_t \sigma_\chi = c_A^2/(\chi V)\int \b E\cdot \b B \,d^3x\,,
\gal
 where $c_A = N_c \sum_f q^2_f e^2/(2\pi^2)$ is the anomaly coefficient, $V$ is the volume of the system and
 $\chi$ is the susceptibility that does not depend on time \cite{Tuchin:2017vwb, Fukushima:2008xe}. Eq.~\eq{a13} can be written in terms of the magnetic helicity defined as
 \ball{a15} 
 \mathcal{H}_\text{em}= \int \b A\cdot \b B\, d^3x\,,
 \gal
 Denoting $\beta = c^2_A/(V\chi)$ yields
\ball{a17}
\beta^{-1} \partial_t  \sigma_\chi =- \partial_t \mathcal{H}_\text{em}-2\sigma_m \mathcal{H}_\text{em}\,.
\gal
Evidently, the total helicity $\mathcal H_\text{tot} = \beta^{-1} \sigma_\chi + \mathcal H_\text{em}$ is not a conserved quantity at finite $\sigma_m$. While the magnetic current is energy non-dissipative, it does dissipate the magnetic helicity. 

\subsection{Meissner effect}\label{sec:meissner}

That the magnetic current does not dissipate energy can also be seen from the fact that under time-reversal $\mathcal{T}$ the current density and magnetic field change signs, implying that the magnetic conductivity $\sigma_m$ is even under $\mathcal{T}$. The same argument indicates that the chiral conductivity $\sigma_\chi$ is also even under $\mathcal{T}$, which, as recently argued by Kharzeev, implies the existence of the ``chiral magnetic superconductivity" \cite{Kharzeev:2016tvd}. 

To see how the supercurrent  induces the Meissner effect,  it is convenient to introduce the ``normal" and ``super" components of the electric field as  
\bal
 \b E_n= -\partial_t \b A\,, \qquad \bm E_s = -\sigma_m \bm A\,.\label{m3}
 \gal
We denote the electric currents induced by each component as
\bal
\b j_n= \sigma_e \b E_n\,,\qquad  \b j_s= \sigma_e \b E_s=-\sigma_e\sigma_m \bm A\,. \label{m6}
\gal
It can be checked that both currents satisfy 
the continuity equation: $\bm\nabla \cdot \bm j_n=\bm\nabla \cdot \bm j_s = 0$. It is straightforward  to see that the super current $\bm j_s$ satisfies the London equations:
\begin{align}
 \bm \nabla \times \bm j_s =\,& -\sigma_e\sigma_m \bm B, \label{m8}\\
\partial_ t \bm j_s =\,& + \sigma_e \sigma_m \bm E_n,\label{m9}
\end{align}
which indicate that $\b j_s$ is indeed a superconducting current. 
The MCSm equations \eq{a3}--\eq{a6} can be rewritten for the pair of fields $\b B$, $\b E_n$ as 
\bal
 & \bm \nabla \cdot \bm E_n = 0, \label{m10}\\
 & \bm \nabla \cdot \bm B = 0,  \label{m11}\\
 -& \bm \nabla \times \bm E_n =\partial_t\bm B,  \label{m12}\\
 & \bm \nabla \times \bm B = \partial_ t \bm E_n + (1+\mathsmaller{\frac{\sigma_m}{\sigma_e}})\bm j_n + \bm j_s + \sigma_\chi \bm B. \label{m13}
\gal
In the stationary limit $\b j_n=0$, $\b E_n=0$ \eq{m8} and \eq{m13} yield
\ball{m20}
 \nabla^2 \bm B = \sigma_e\sigma_m \bm B - \sigma_\chi \bm \nabla\times \bm B\,,
\gal
which can also be seen directly from \eq{a9}. The super component of the electric field satisfies the same equation. Indeed, taking the Laplacian of the second equation in \eq{m3} and using \eq{m13} we obtain
\ball{m21}
 \nabla^2 \bm E_s = \sigma_e\sigma_m \bm E_s - \sigma_\chi \bm \nabla\times \bm E_s\,.
\gal
In the anomaly-free case $\sigma_\chi=0$, Eqs.~\eq{m20},\eq{m21} imply that the electromagnetic field decays exponentially inside the conductor over the London penetration length  $\ell=1/\sqrt{\sigma_e\sigma_m}$.  In the ideal conductor limit $\ell\to 0$, the field is expelled, which is the Meissner effect. 

\begin{figure}
 \centering 
 \includegraphics[width=0.4\textwidth]{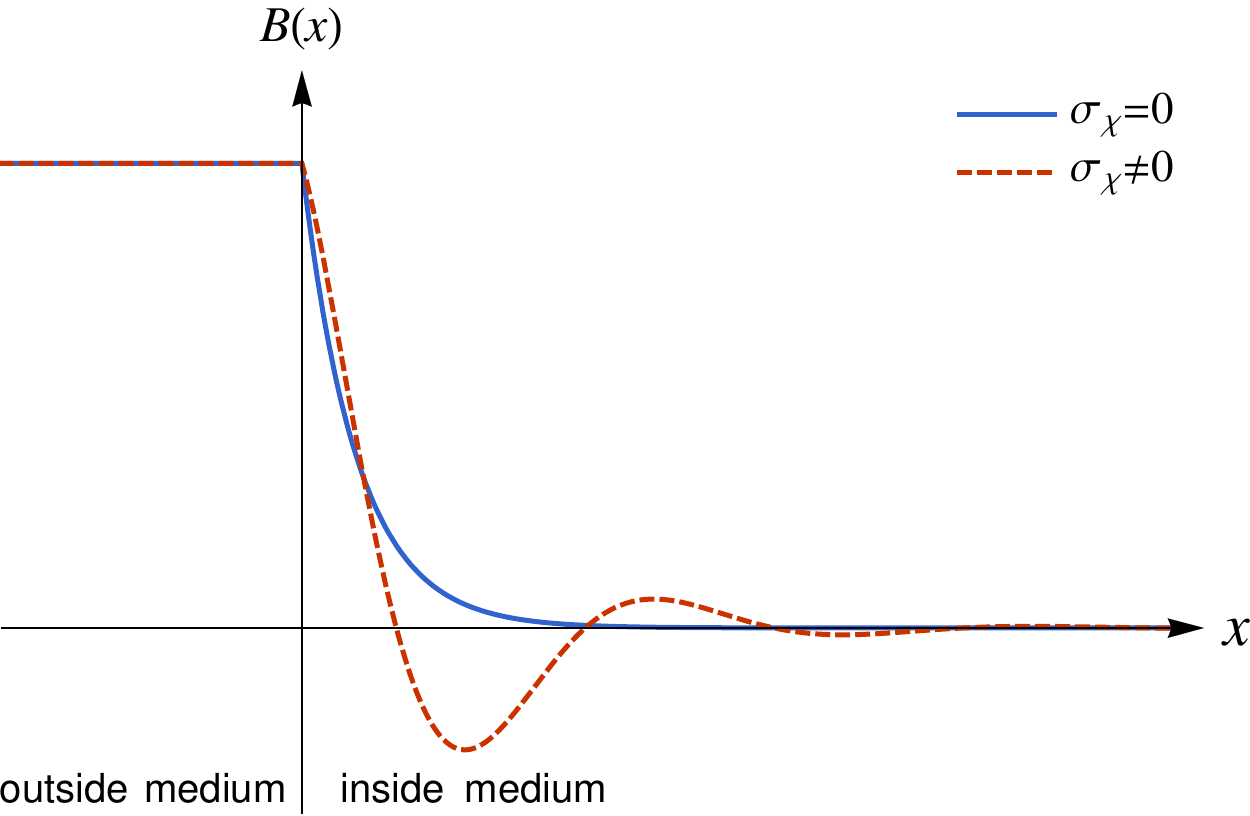}
 \caption{Meissner effect in a chiral medium.}
 \label{fig:Meissner}
\end{figure}

To analyze \eq{m20} and \eq{m21} at finite constant  $\sigma_\chi$, we expand $\b B$ and $\b E_s$ into a complete set of eigenfunctions $\b W_{\b k\lambda}(\b x)$ of the curl operator, known as the Chandrasekhar-Kendall (CK) states  \cite{CK}. Here $\b k$ labels the Laplacian eigenvalues, in particular $k\ge 0$ is the wavenumber, and $\lambda=\pm 1$ is helicity. 
Using 
\ball{a18-1}
\b\nabla\times \b W_{\b k\lambda}(\b x)= \lambda k\b W_{\b k\lambda}(\b x)\,,
\gal
we find that the wavenumber $k$ satisfies
\ball{m25}
 k = \frac{\lambda\sigma_\chi}{2} \pm \sqrt{\frac{1}{4}\sigma^2_\chi-\sigma_e\sigma_m}.
\gal
Since the CK states oscillate at large $x$, we observe that the electric and magnetic fields exponentially decay in matter if  $\sigma_\chi^2 < 4\sigma_e\sigma_m$. The corresponding London  penetration length is
\ball{m30}
\ell = 1\bigg/\sqrt{\sigma_e\sigma_m - \frac{1}{4}\sigma^2_\chi}\,.
\gal 
Additionally, in the chiral medium, the electric and magnetic fields  oscillate as they decay,  see \fig{fig:Meissner}.  At $\sigma_\chi^2 \ge 4\sigma_e \sigma_m$ there is no  Meissner effect. In fact, as we will argue in \sec{sec:d}, at such values of $\sigma_\chi$, the helicity of magnetic field is unstable, growing exponentially in time.  

Thus far, when discussing the Meissner effect, we ignored  the time-dependence of  $\sigma_\chi$ which stems from the chiral anomaly equation \eq{a17}. Generally, it can be expected that during the chiral evolution, the medium may go through both the superconducting and the normal phase. However, as we will argue, the fixed point of the chiral evolution is superconducting.

\subsection{Late-time dynamics}\label{sec:ca}

Equations~\eq{a9} and \eq{a17}  describe the time-evolution of the magnetic helicity and the chiral conductivity. Although their time-evolution depends  on the initial conditions, we can make a number of general statements about the late-time behavior of the system. We know from the previous studies that in a chiral medium without magnetic charges, i.e.\ $\sigma_m=0$, all helicity is eventually transferred from the medium to the magnetic field so that in the final state at $t\to \infty$ the chiral conductivity vanishes, while the magnetic helicity is maximal \cite{Boyarsky:2011uy,Hirono:2015rla}. At finite $\sigma_m$, helicity is dissipated by the magnetic currents so that the only possible finial state with constant $\sigma_\chi$ has $\mathcal{H}_\text{em}=0$. However, since the total helicity  is not conserved at finite $\sigma_m$, it does not restrict a possible asymptotic value of $\sigma_\chi$. The system can thus settle to a stationary state with finite chiral conductivity and vanishing magnetic helicity\footnote{In the static case, when the chiral conductivity and magnetic helicity are time-independent, the only possible solution to \eq{a17} is $\mathcal{H}_\text{em}\equiv 0$.}. 

We can refine our conclusion by studying how the finial state is achieved. According to the helicity balance equation \eq{a17}, the magnetic helicity decreases at later times as $\mathcal{H}_\text{em}\sim e^{-2\mu t}$ while the chiral conductivity decreases as $\sigma_\chi \sim \sigma_\infty + \mathcal{O}(e^{-2\mu t})$, where $\mu$ and $\sigma_\infty$ are positive constants, which depend on the parameters of the system and the initial conditions.   In view of \eq{a15}, the late time behavior of the magnetic field is $B\sim e^{-\mu t}$.\footnote{In the absence of magnetic current, the magnetic field decays at later times as a power law \cite{Hirono:2015rla,Tuchin:2017vwb}.} This motivates us seeking for an asymptotic solution to \eq{a9} in the form $\b B(\b x, t) = \b B'(\b x)e^{-\mu t}$ which yields 
\ball{a19}
-\nabla^2 \b B'+(\sigma_e-\mu)(\sigma_m-\mu)\b B'-\sigma_\infty \b \nabla\times \b B'=0\,.
\gal
This equation shares a complete set of eigenfunctions $\b W_{\b k\lambda}(\b x)$ with the curl operator \cite{CK}.  Expanding the magnetic field in this basis and using \eq{a18-1} we find that 
\ball{a21}
k=k_\pm = \lambda\sigma_\infty/2\pm \sqrt{\sigma_\infty^2/4-(\sigma_e-\mu)(\sigma_m-\mu)}\,.
\gal
Thus, the real solutions to \eq{a19} exist only if $\sigma_\infty$ satisfies
\ball{a23}
\sigma_\infty^2\ge 4 (\sigma_e-\mu)(\sigma_m-\mu)\,.
\gal
In a medium without magnetic monopoles $\sigma_m=0$, the minimum value of $k_\pm$ vanishes indicating that the helicity and energy can be transferred from the medium to the infrared modes of the magnetic field $k\to 0$ in the process known as the inverse cascade \cite{Boyarsky:2011uy}. In contrast, in a medium with magnetic charges, $k_\pm$ do not vanish at a finite $\sigma_\infty$, implying that the inverse cascade terminates at finite chiral conductivity, while the magnetic field and magnetic helicity exponentially decay. 

Now as we got a glimpse into the properties of the chiral medium with magnetic monopoles, we turn to a more quantitative discussion.

\section{Adiabatic approximation}\label{sec:d}
 
 In view of the analysis in the preceding section it is advantageous to  proceed by expanding the vector potential into the complete set of the CK states
\ball{b2}
\b A = \sum_{\b k, \lambda}\left[a_{\b k\lambda}(t)\b W_{\b k\lambda}(\b x)+ a^*_{\b k\lambda}(t)\b W^*_{\b k\lambda}(\b x)\right]\,.
\gal
In particular, in the Cartesian coordinates it is a set of  the circularly polarized plane waves 
\ball{b4} 
\b W_{\b k\lambda}(\b x)= \frac{\b\epsilon^\lambda}{\sqrt{2kV}}e^{i\b k\cdot \b x}\,,
\gal
which are eigenstates of the curl operator  with eigenvalues $\lambda k$, $\b\epsilon^\lambda$ is the polarization vector with $\lambda= +1 (-1)$ 
corresponding to the right-handed (left-handed) polarization and $V$ is volume. Substituting \eq{b2} into \eq{a9}   one derives an equation 
\ball{b6}
k^2a_{\b k \lambda} + \ddot a_{\b k \lambda} = - (\sigma_e+\sigma_m) \dot a_{\b k \lambda}-\sigma_e\sigma_m a_{\b k \lambda} + \lambda k \sigma_\chi(t) \, a_{\b k \lambda}\,.
\gal
The magnetic helicity  \eq{a15} can be written as 
\ball{b8}
\mathcal{H}_\text{em}= \int \b A\cdot \b B\, d^3x = \sum_{\b k,\lambda}\lambda |a_{\b k\lambda}|^2\,.
\gal
Upon substitution into \eq{a17}, it yields an implicit equation for $\sigma_\chi(t)$.

\begin{figure}[ht]
\begin{tabular}{lcr}
     \includegraphics[height=7cm]{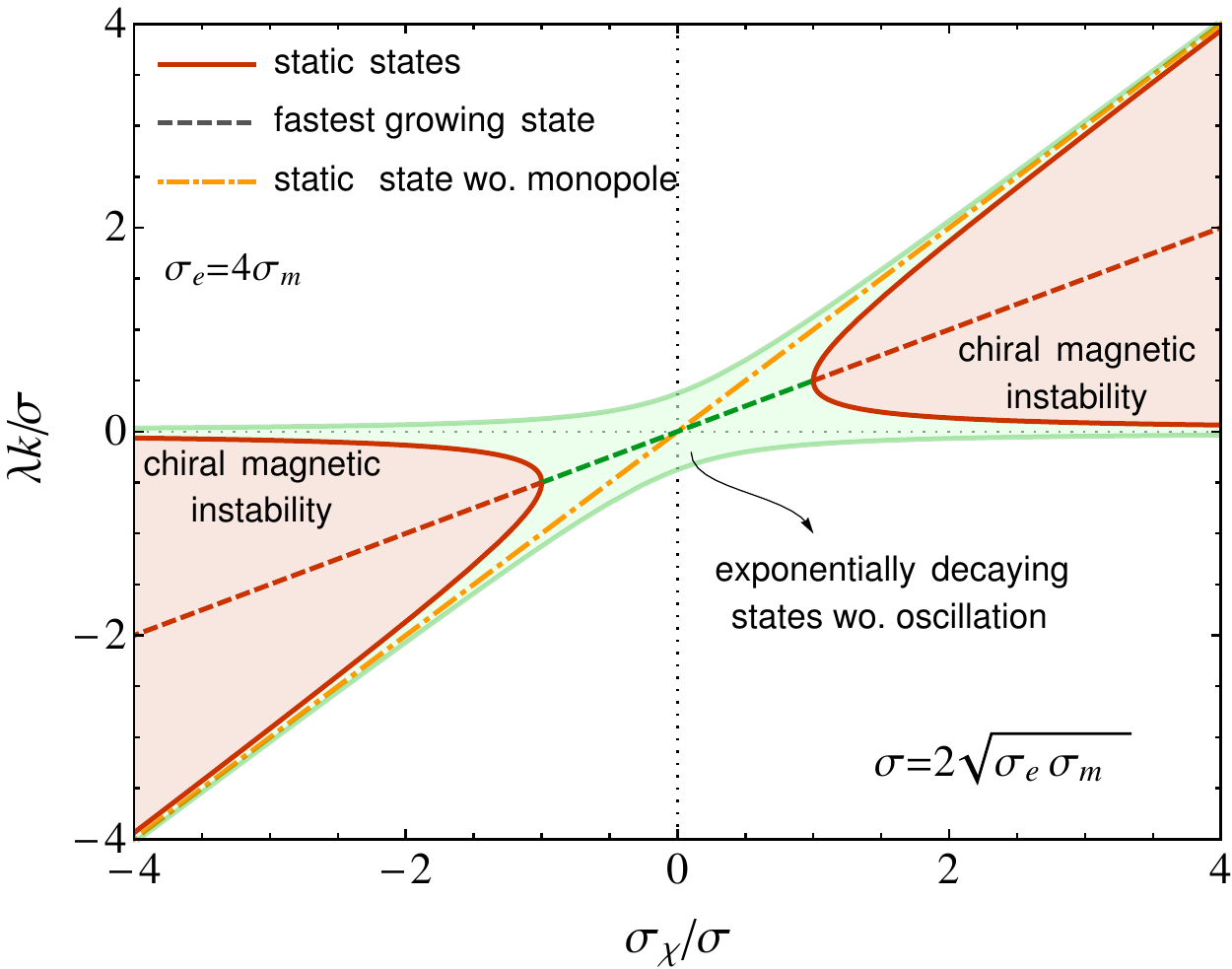}
     \end{tabular}
 \caption{Properties of the dispersion relation \eq{d5}. The upper (lower) half corresponds to 
 states with $\lambda = +1$ ($\lambda = -1$).
 Regions with chiral magnetic instability where the Meissner phase is not possible  are shown in red.  They are enclosed by the static CK states. The green (white) region consists of exponentially decaying
 CK states without (with) oscillation.  The border  lines for oscillation shown here correspond to $\sigma_e = 4 \sigma_m$. 
We have also defined $\sigma=2\sqrt{\sigma_e\sigma_m}$. }
\label{fig5}
\end{figure}

A more detailed analysis of the time-evolution problem can be done using the adiabatic approximation. It also allows one to consider media 
with realistic values of electric conductivity.
The adiabatic approximation consists in writing the amplitudes in the form 
\ball{d1}
a_{\b k \lambda}= a_{\b k\lambda}(0) e^{-i\int_0^t\omega_{k\lambda}(t')dt'}
\gal
and assuming that $\omega_{k\lambda}(t)$ is a slow varying function. Substituting \eq{d1} into \eq{b6} and neglecting terms proportional to $\dot \omega_{k\lambda}$, one obtains a quadratic equation 
\ball{d3}
k^2-\omega_{k\lambda}^2= i\omega_{k\lambda}(\sigma_e+\sigma_m)-\sigma_e\sigma_m + \lambda k \sigma_\chi
\gal
that has two solutions 
\ball{d5}
\omega_{k\lambda}(t)= -\frac{i(\sigma_e+\sigma_m)}{2} + \lambda_1\frac{i}{2}\sqrt{(\sigma_e+\sigma_m)^2 +4(\sigma_\chi \lambda k-\sigma_e\sigma_m-k^2)}\,,
\gal
where $\lambda_1=\pm 1$. 

The dispersion relation \eq{d5} has the following properties summarized on the diagram \fig{fig5}:
\begin{enumerate}
\item Modes with $\lambda_1>0$ and $\sigma_\chi \lambda k-\sigma_e\sigma_m-k^2\ge 0$ are growing, i.e.\ unstable, because $\text{Im}\,\omega_{k\lambda}>0$. This means that the magnetic field and the corresponding magnetic helicity grow exponentially with time through the transfer of helicity from the medium. The corresponding momentum values are 
\ball{f1}
\frac{\sigma_\chi}{2} - \frac{1}{2}\sqrt{\sigma^2_\chi - 4\sigma_e\sigma_m} \le k \le \frac{\sigma_\chi}{2} + \frac{1}{2}\sqrt{\sigma^2_\chi
- 4\sigma_e\sigma_m}\,.
\gal
This kinematic region exists only if $\sigma_\chi ^2> 4 \sigma_e\sigma_m$. Eventually, after a long time, the magnetic helicity vanishes (as explained in \sec{sec:ca}) while the chiral conductivity settles into a stationary state with  $\sigma_\chi^2\to \sigma_\infty^2 \le 4\sigma_e\sigma_m$, which is the value of the chiral conductivity at which the inverse cascade terminates. Termination of the inverse cascade is a distinctive feature of the MCSm theory. Without monopole, the inverse cascade is self-similar \cite{Hirono:2015rla} and 
is only terminated as the characteristic instability wavelength $\lambda_\star \sim 2/\sigma_\chi$ grows larger than the size of the system.

\item The CK modes in the region $\sigma_\chi^2 \le 4\sigma_e \sigma_m$ are stable; this is the Meissner phase. The
magnetic field and magnetic helicity decay exponentially with ($\mathrm{Re}\, \omega_{\lambda, k}\ne0$) or without oscillation ($\mathrm{Re}\, \omega_{\lambda, k}=0$) depending on 
the values of $k$, $\sigma_e$ and $\sigma_m$. It is seen in \fig{fig5} that modes $k\le \sigma_\chi$  ($\lambda=+1$) are always non-oscillating. The growing and damped modes are separated by the static CK states  with $\omega_{k\lambda}=0$.

\end{enumerate}

\section{Fastest Growing State (FGS) model}\label{sec:d+}

In order to better understand the time-dynamics of the MCSm system it is useful to use a model that on the one hand, has all properties discussed in the previous sections, while on the other hand, is analytically solvable and hence easy to interpret. In the absence of the magnetic current $\b j_m=0$, such a model, dubbed the Fastest Growing State model, was developed by one of us in \cite{Tuchin:2017vwb}. It reproduces the essential features of the time evolution found in numerical calculations and provides a number of novel insights. In this and the following sections we generalize this model to include the magnetic current. We will see, however, that its applicability is restricted to the case $\sigma_e\ge\sigma_m$.

Time evolution at later times is determined by the modes with negative imaginary part of $\omega_{k\lambda}$. Among them there is the 
fastest growing mode $k_\star$ such that $\dot \omega_{k_\star\lambda}=0$. Taking the time derivative of
\eq{d3} one finds that the fastest growing mode has the momentum
\ball{d7}
k_\star= \frac{\sigma_\chi \lambda}{2}\,,
\gal
which is independent of the electric and magnetic conductivities. Using this in \eq{d5} one finds the amplitude of the fastest growing mode 
\ball{d9}
a_\star(t) = a_\star(0) e^{\frac{1}{2}\gamma(t)}\,,
\gal
where 
\ball{d11}
\gamma(t) = \int_0^t \left[ \sqrt{(\sigma_e-\sigma_m)^2 +\sigma_\chi^2(t')} - (\sigma_e+\sigma_m)\right] dt'\,.
\gal

At later time one can approximate the sum in \eq{b8} by the fastest growing amplitude \eq{d9}.  For definitiveness we also assume that $\sigma_\chi$ is positive implying that $\lambda=+1$. Thus, the magnetic
helicity becomes
\ball{d13}
\mathcal{H}_\text{em}(t) = f \mathcal{H}_\text{tot}(0) e^{\gamma(t)}\,,
\gal
where $f = \mathcal H_\text{em}(0)/\mathcal H_\text{tot}(0) \ge 0$ is the fraction of the total helicity in magnetic field at
$t=0$.

It is convenient to define the dimensionless conductivities $\sigma_\chi \to \sigma_\chi/\alpha$, $\sigma_m \to \sigma_m/\alpha$, 
$\sigma_e \to \sigma_e/\alpha$ and dimensionless time $t\to \alpha t$, where $\alpha = \beta \mathcal H_\text{tot}(0) =
\mathcal H_\text{tot}(0) c_A^2/(\chi V)$ is a characteristic energy scale.
Using these notations, as well as \eq{d13}, we can write \eq{a17} as
\ball{d17}
\partial_t  \sigma_\chi =-f(\dot \gamma +2\sigma_m) e^\gamma  \,.
\gal
Let us now divide this equation by $d\gamma/dt$ from \eq{d11}. We have
\ball{d19}
 \frac{d\sigma_\chi}{d\gamma}= -\frac{ \sqrt{(\sigma_e-\sigma_m)^2 +\sigma_\chi^2 } -\sigma_e+\sigma_m}{\sqrt{(\sigma_e-\sigma_m)^2 +\sigma_\chi^2 } -\sigma_e-\sigma_m}fe^\gamma\,.
\gal
Considering the chiral conductivity to be a function of $\gamma$ this equation can be easily integrated. The solution is 
\ball{d21}
\gamma= \ln \left\{ 1-f^{-1}\left[ F(\sigma_\chi)-F(1-f)\right]\right\}\,,
\gal
where we defined
\bal
F(\sigma_\chi)= \frac{1}{\sigma_\chi}\left\{ \sigma_\chi^2+2\sigma_m\left[\sqrt{\sigma_\chi^2+(\sigma_e-\sigma_m)^2}+ \sigma_e-\sigma_m\right]\right. \nonumber\\
\left.
-2\sigma_\chi \sigma_m\ln\left[ \sigma_\chi +\sqrt{\sigma_\chi^2+(\sigma_e-\sigma_m)^2}\right]\right\}\,,
\label{d23}
\gal
and used $\sigma_\chi (0)=1-f$. In the limit $\sigma_m\to 0$, $F\to  \sigma_\chi$. 
Substituting \eq{d21} into \eq{d17},  we derive the equation that governs the time evolution of the chiral conductivity
\ball{d33}
\dot \sigma_\chi= -\left[f+F(1-f)-F(\sigma_\chi)\right]\left(\sqrt{(\sigma_e-\sigma_m)^2+\sigma^2_\chi}- \sigma_e+\sigma_m\right) \,. 
\gal
This is the main equation of the FGS model. Once {\eq{d33}} is solved, one can compute $\gamma$ using {\eq{d21}} and magnetic helicity using {\eq{d13}}.

Since the right-hand-side of \eq{d33} is negative, the chiral conductivity  is a monotonically decreasing function. At later times it approaches a stationary solution $\sigma_\infty$. In general, the stationary solution $\sigma_\infty$ is non-zero, in contrast to the case
without magnetic monopoles monopole ($\sigma_m=0$). Moreover, $\sigma_\chi=0$ is a stationary solution only if $\sigma_e\ge\sigma_m$. Indeed, in this case the right-hand-side of \eq{d33} vanishes. If  $\sigma_e<\sigma_m$, the chiral conductivity can become negative indicating the breakdown of the model \footnote{The chiral conductivity can change sign and become negative as indicated in e.g.\ \eq{a23}. However,  the FGS model is not suitable for such analysis. }. From now on we concentrate on the $\sigma_e\ge\sigma_m$ case.

Once \eq{d33} is solved, the magnetic field can be computed as
\ball{e1}
B = B_0\sqrt{{\sigma_\chi }/{(1-f)}}\, e^{\frac{1}{2}\gamma}\,.
\gal
Clearly it exponentially decays at long times, as $\gamma<0$ and $\sigma_\chi\to\sigma_\infty$ as $t\to\infty$ [cf. Eqs.~(\ref{d21}--\ref{d33})]. The magnetic helicity reads using \eq{d13} and \eq{d21}
\ball{e4}
\mathcal H_\text{em} = \mathcal H_\text{tot}(0) \left[ f + F(1-f) - F(\sigma_\chi)\right].
\gal
It can be shown that $F'(\pm2\sqrt{\sigma_e\sigma_m}) = 0$. Therefore, $\mathcal H_\text{em}$ 
peaks at $t_\text{pk}$ defined as $\sigma_\chi(t_\text{pk}) = 2\sqrt{\sigma_e\sigma_m}$. 
The total helicity, however, always decreases in presence of $\sigma_m$.

\begin{figure}
 \centering 
\subfloat[\ $f=0.5$, $\sigma_e = 10 \sigma_m$]{
  \includegraphics[width=0.33\textwidth]{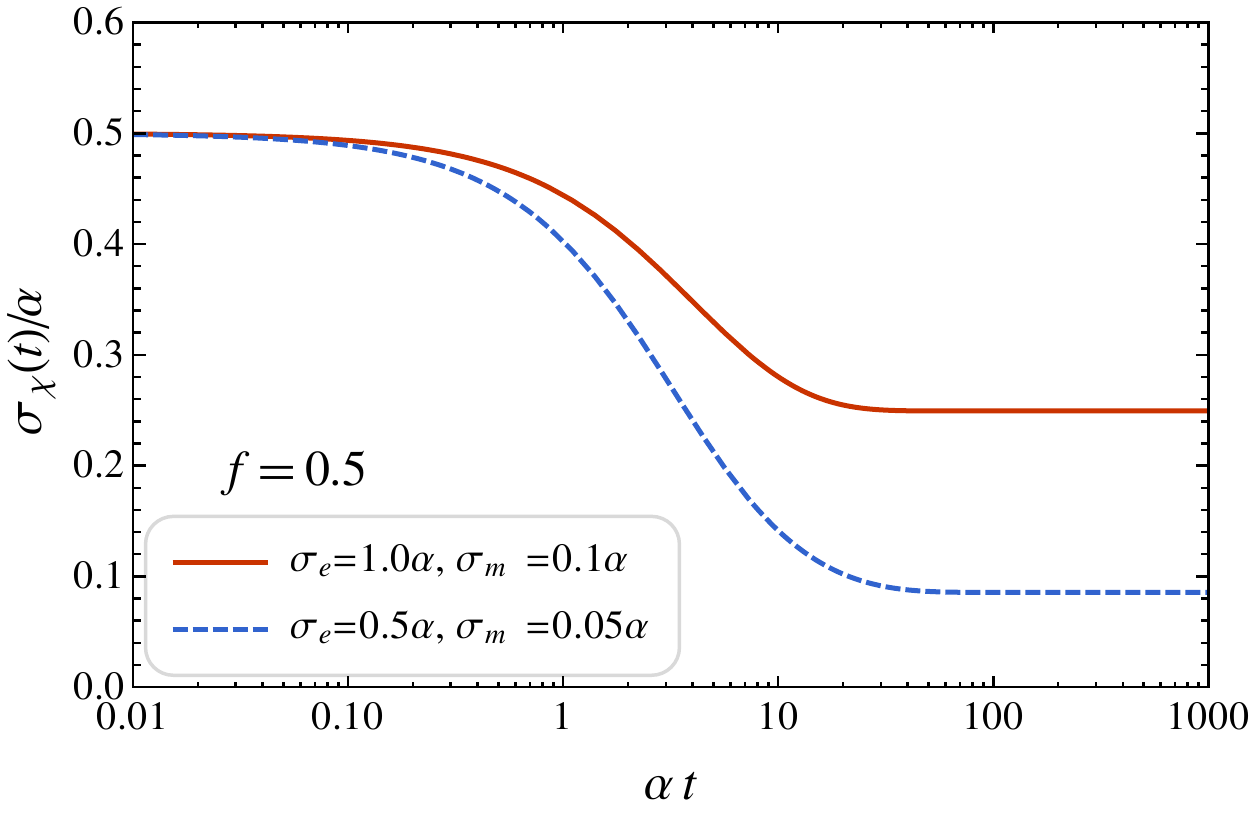}
  \includegraphics[width=0.33\textwidth]{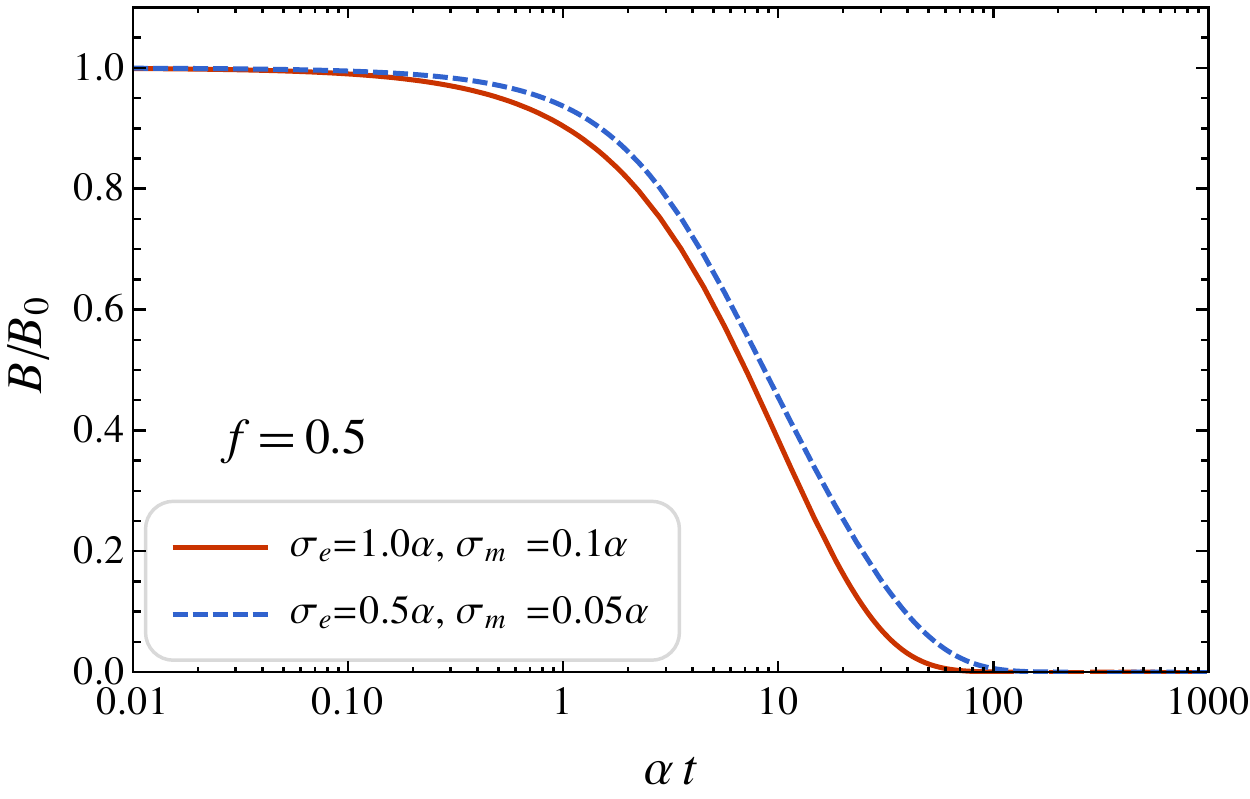}
  \includegraphics[width=0.33\textwidth]{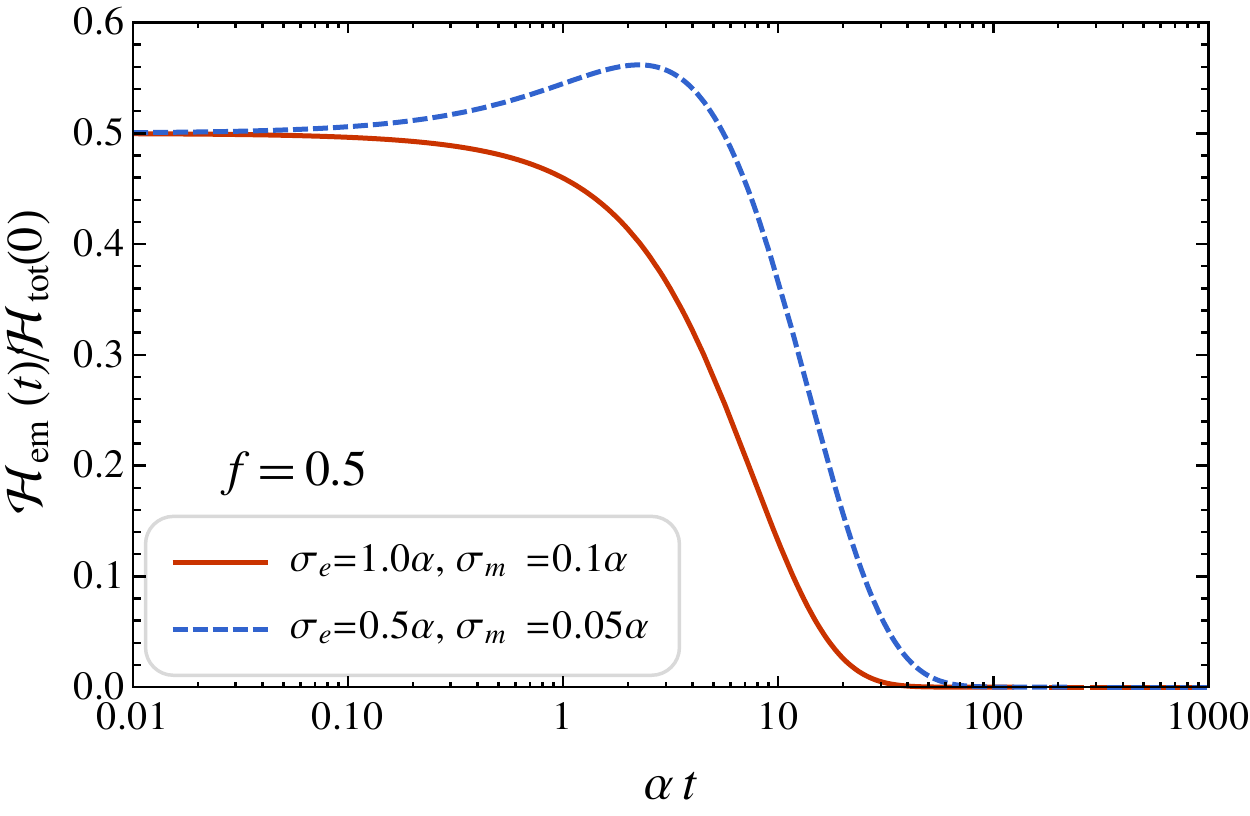}
}

\subfloat[\ $f=10^{-4}$, $\sigma_e = 10 \sigma_m$]{
  \includegraphics[width=0.33\textwidth]{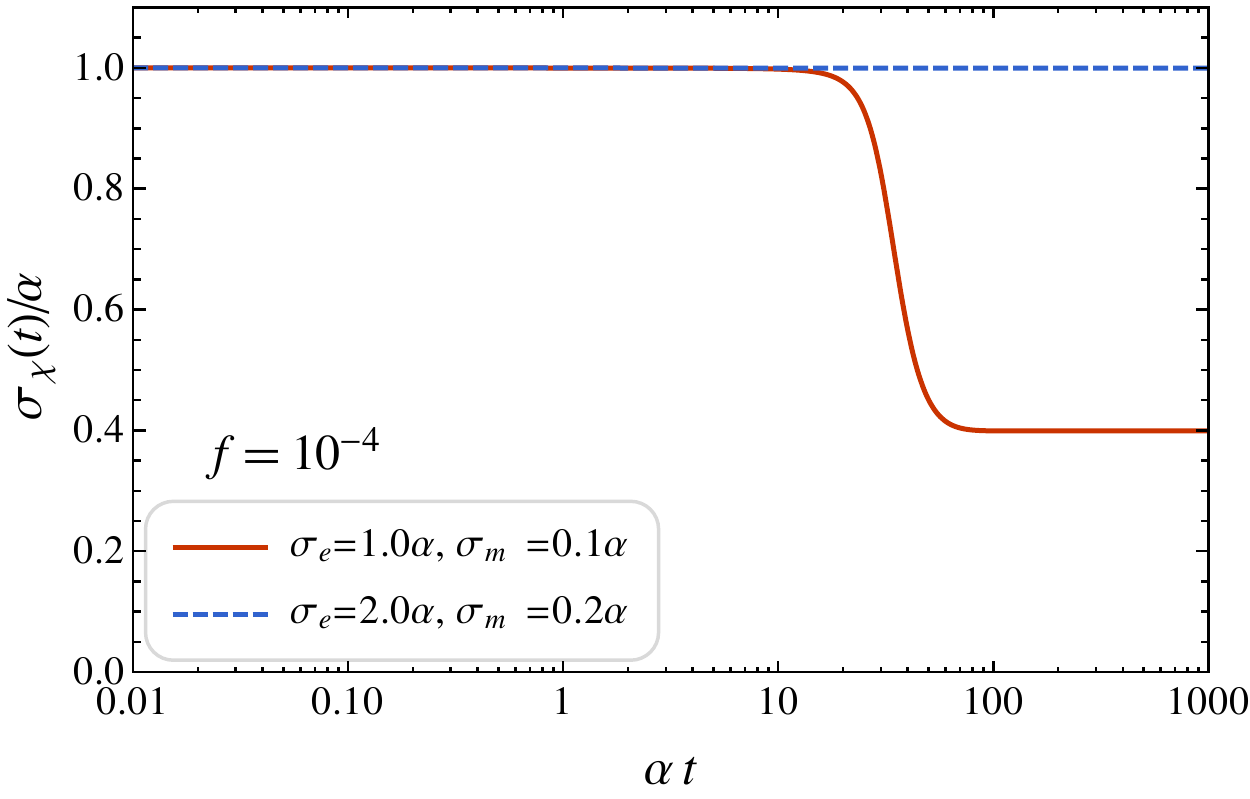}
  \includegraphics[width=0.33\textwidth]{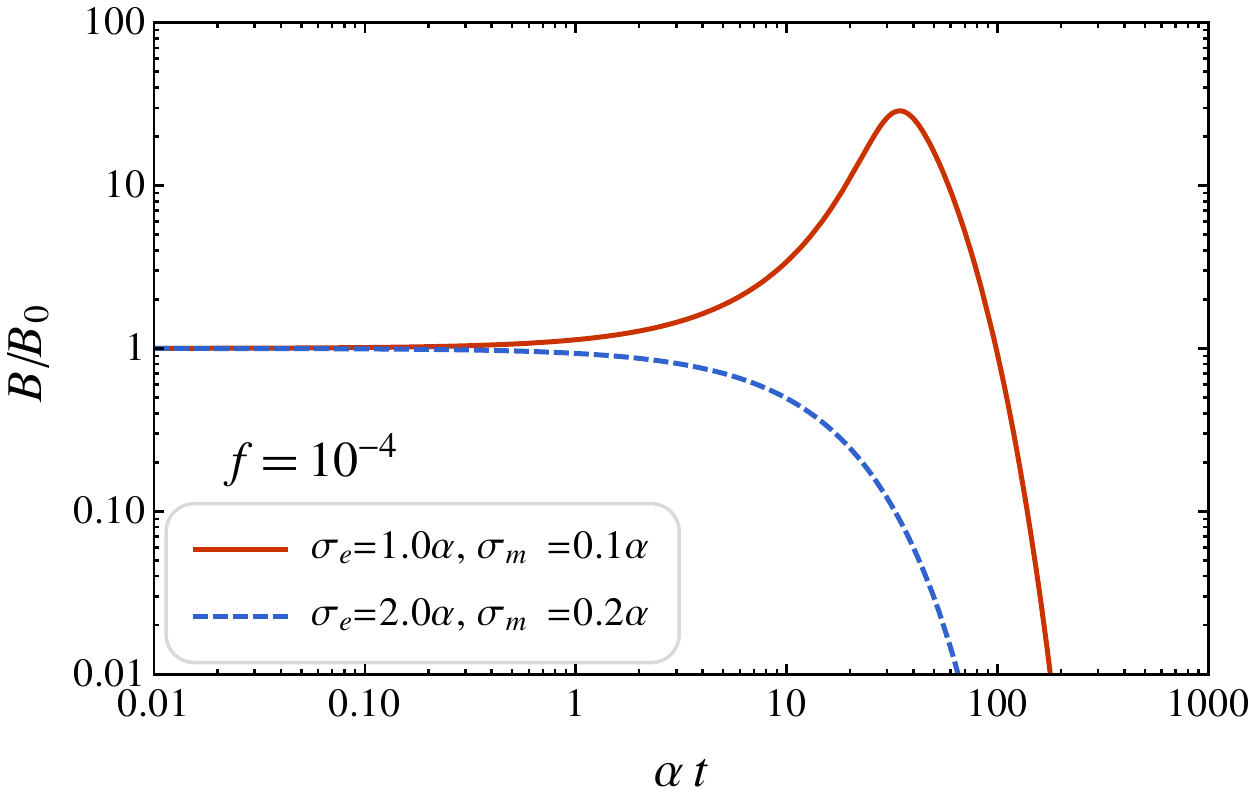}
  \includegraphics[width=0.33\textwidth]{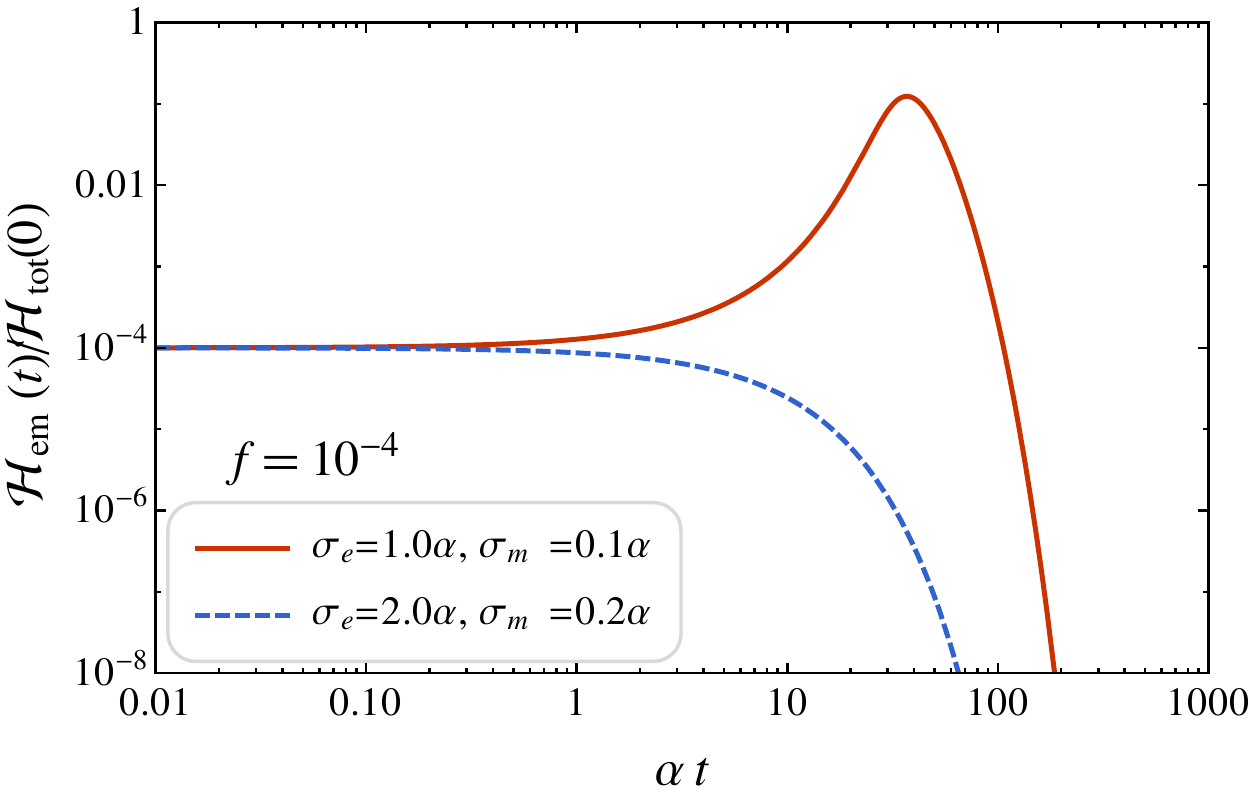}
}

\subfloat[\ $\sigma_e = \alpha, \sigma_m = 0.1\alpha$]{
  \includegraphics[width=0.33\textwidth]{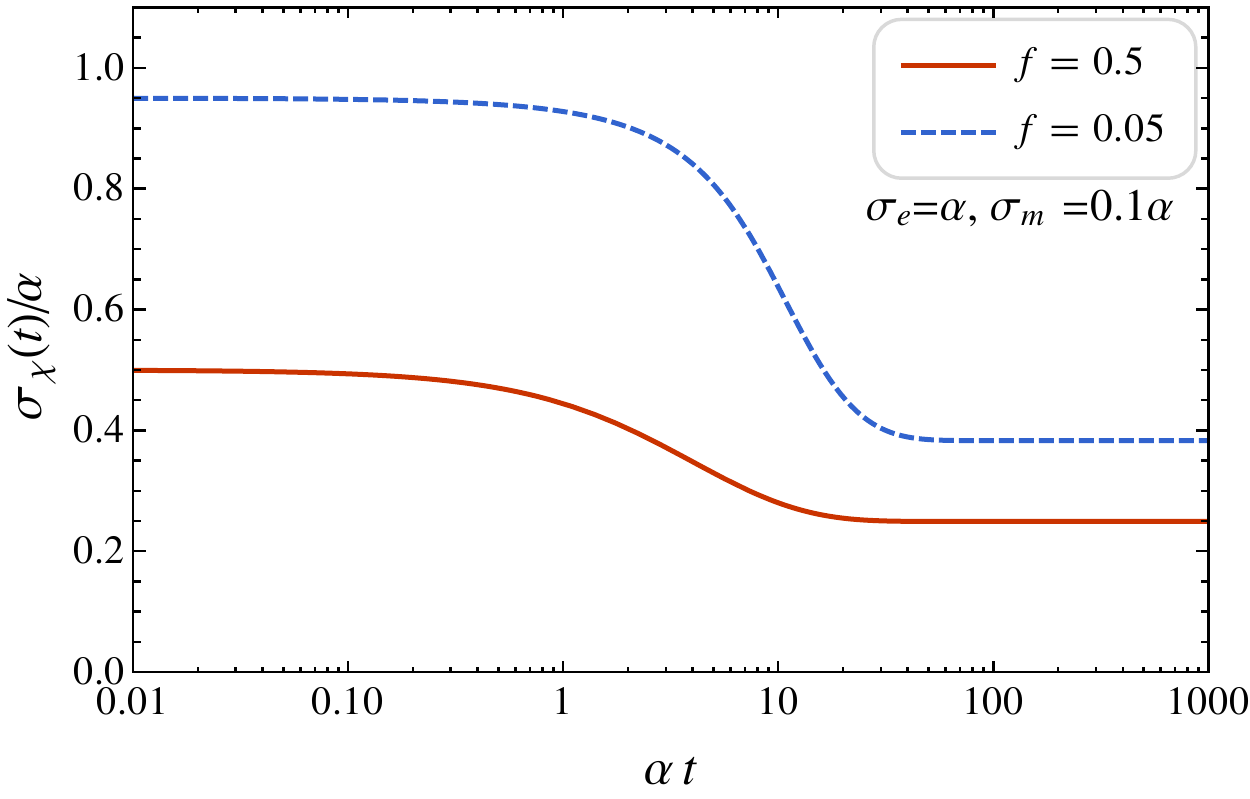}
  \includegraphics[width=0.33\textwidth]{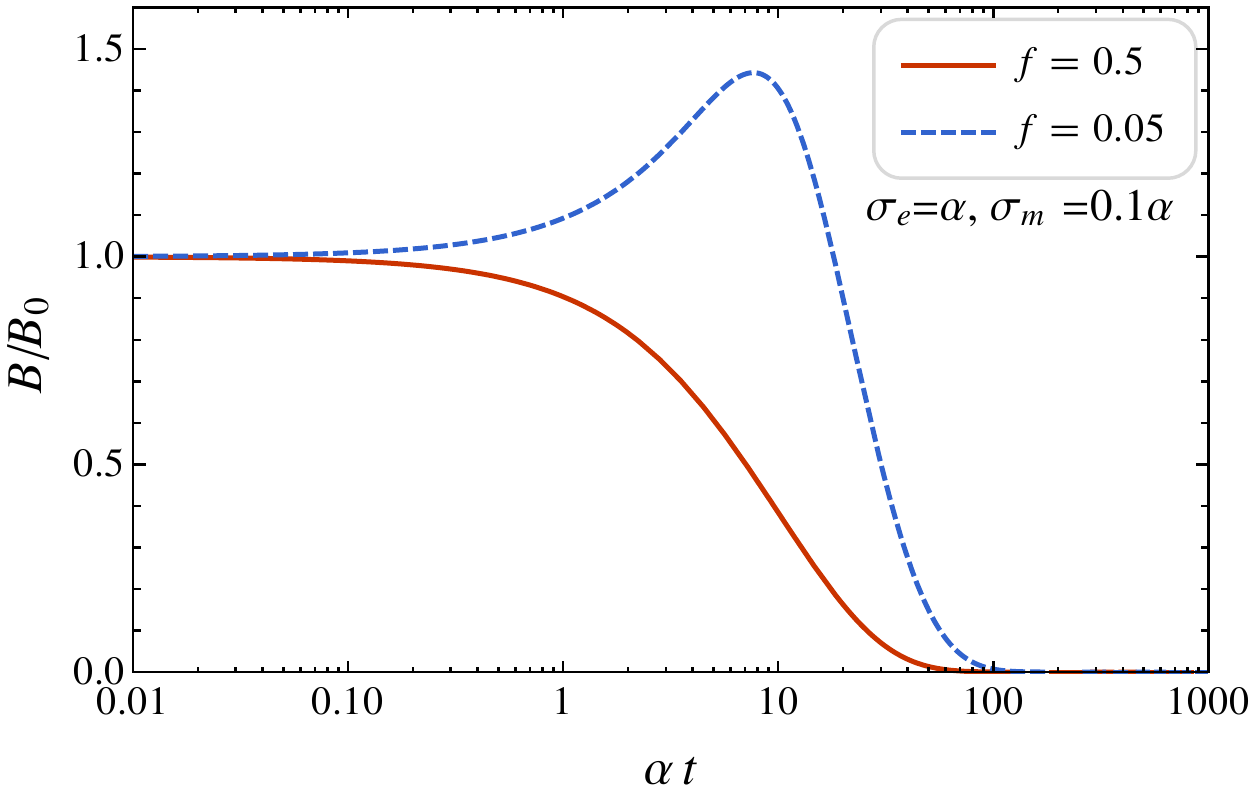}
  \includegraphics[width=0.33\textwidth]{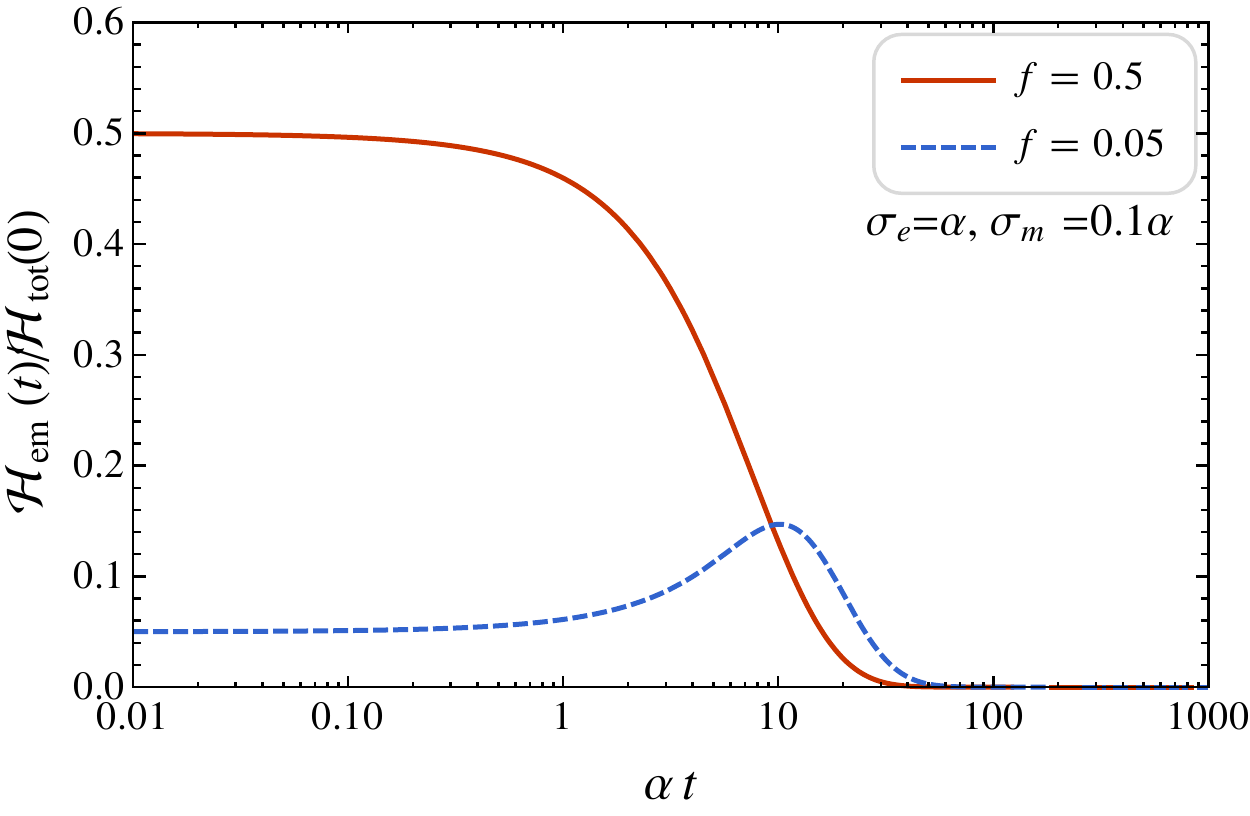}
}
\caption{Left column: chiral conductivity $\sigma_\chi(t)$ as a function of time for different initial conditions $\sigma_\chi(0)=1-f$ and different electric and magnetic conductivities. Center column: the corresponding evolution of the magnetic field. Right column: the corresponding evolution of the magnetic conductivity. $\alpha = \mathcal H_\text{tot}(0) c_A^2/(\chi V)$ is a characteristic energy scale.  }
\label{fig:solutions}
\end{figure}

\begin{figure}[ht]
\centering 
 \includegraphics[width=0.45\textwidth]{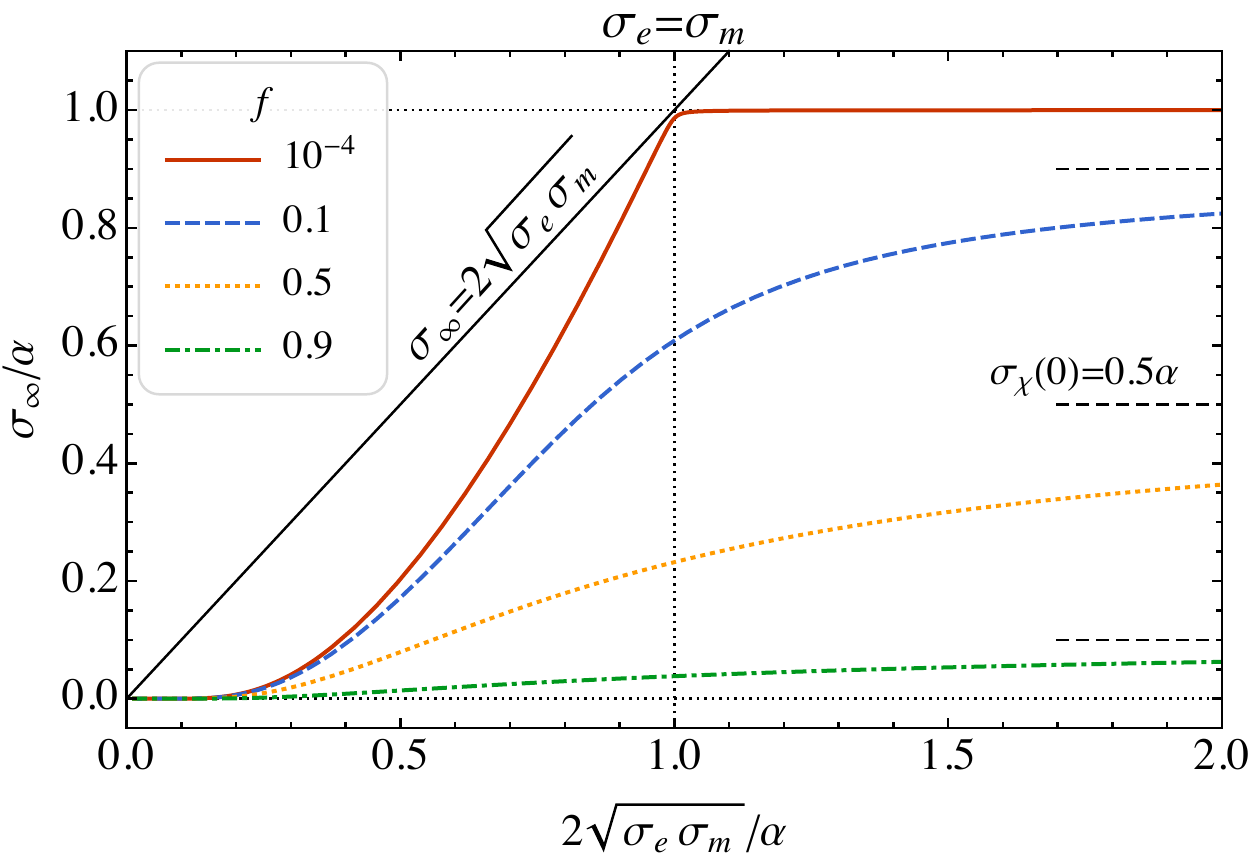}
 \caption{Asymptotic values of the chiral conductivity $\sigma_\infty$ at different initial conditions $\sigma_\chi (0)=1-f$. The diagonal line separates the stable region below it from the unstable one above it. }
\label{fig1}
\end{figure}

Representative solutions of \eq{d33} are shown in Fig.~\ref{fig:solutions}. It is seen that at $t\to \infty$ the value of the chiral conductivity approaches a constant that we labeled as $\sigma_\infty$ in \sec{sec:ca}. To discuss the possible values of the chiral conductivity at $t\to \infty$ it is convenient to write \eq{d33} using \eq{e4} as 
\ball{e6}
\mathcal H_\text{tot}(0)\,\dot \sigma_\chi= -\mathcal H_\text{em}(\sigma_\chi) \left(\sqrt{(\sigma_e-\sigma_m)^2+\sigma^2_\chi}- \sigma_e+\sigma_m\right) \,. 
\gal
Here $\mathcal H_\text{tot}(0)$ is the value of the total helicity at $t=0$, whereas $\mathcal H_\text{em}(\sigma_\chi)$ is the magnetic helicity as a function of $\sigma_\chi$. The stationary solutions that the chiral conductivity approaches as $t\to \infty$ satisfy $\dot\sigma_\chi = 0$. Eq.~\eq{e6} always admits a stationary solution  $\sigma_\infty =0$  due to the vanishing of the expression in the round brackets as $\sigma_\chi\to 0$ (since $\sigma_e\ge\sigma_m$). The remaining stationary solutions are the real positive\footnote{As per assumption below Eq.~\eq{d11}.} roots of the equation $\mathcal H_\text{em}(\sigma_\chi) =0$. Since this equation always has non-trivial roots, the trivial stationary state is never reached, see \fig{fig1}. 

%
For a given initial condition $\sigma_\chi(0)=1-f$, the chiral conductivity settles to the largest root that satisfies $\sigma_\infty\le \sigma_\chi(0)$. 
We can derive a universal bound $\sigma_\infty \le 2\sqrt{\sigma_e\sigma_m}$. It is trivially satisfied if $\sigma_\chi(0) \le 2\sqrt{\sigma_e\sigma_m}$. 
If $\sigma_\chi(0) > 2\sqrt{\sigma_e\sigma_m}$, noting that $\sigma_\chi = 2\sqrt{\sigma_e\sigma_m}$ is a maximum of $\mathcal H_\text{em}(\sigma_\chi)$, we have $\mathcal H_\text{em}(2\sqrt{\sigma_e\sigma_m}) \ge \mathcal H_\text{em}(\sigma_\chi(0)) = f \mathcal H_\text{tot} \ge 0$. Since $\sigma_\chi = 2\sqrt{\sigma_e\sigma_m}$ is the only maximum in the interval $\big[0,  \sigma_\chi(0)\big]$, these two inequalities imply
that the largest root is $\sigma_\infty \le 2\sqrt{\sigma_e\sigma_m}$. This relation can also be seen in the numerical solutions in \fig{fig1}.

\section{Stability of magnetic helicity and magnetic field}\label{sec:e}

It is seen in \fig{fig:solutions} that during the initial stage of the evolution, the magnetic helicity can either grow or decay with time.  In the former case we say that the magnetic helicity is unstable whereas in the later case it is stable. Our goal in this section is to derive the stability condition. The main result is given by \eq{e8}.

\medskip 

We can derive the stability condition by requiring that magnetic helicity be always decreasing function of time, viz.\ $\dot{\mathcal{H}}_\text{em}<0$. It then follows from \eq{e4} that $F'(\sigma_\chi)\dot \sigma_\chi>0$. 
Since $\dot \sigma_\chi <0$ (see \eq{e6}) we conclude that $F'(\sigma_\chi)<0$ for any $\sigma_\chi$. This is the same condition as \eq{f1} from the more general analysis of the dispersion relation, which shows that FGS captures the main feature of the theory. Using \eq{d23} we obtain $\sigma_\chi^2 \le 4 \sigma_e \sigma_m$ for any $\sigma_\chi$. Finally, since $\sigma_\chi\le \sigma_\chi(0)$,  we derive the magnetic helicity is stable (meaning monotonically decreasing) if the chiral evolution starts from the initial condition $\sigma_\chi(0)=1-f$ 
\ball{e8}
\sigma_\chi^2(0) \le 4 \sigma_e \sigma_m\,.
\gal
Conversely, the magnetic helicity is unstable if  $\sigma_\chi^2 > 4 \sigma_e \sigma_m$ for any $\sigma_\chi$. Observing that $\sigma_\chi\ge \sigma_\infty$, we conclude that the magnetic helicity is unstable if $\sigma_\infty^2 > 4 \sigma_e \sigma_m$. We see in \fig{fig1} that this condition is never satisfied. Therefore Eq.~\eq{e8} is the only non-trivial stability condition. As we argued in \sec{sec:meissner}, if it is satisfied, the magnetic field is expelled from the medium.

In particular case $f=0$, i.e.\ no magnetic field at $t=0$, there exists a static solution $\sigma_\chi(t) = \sigma_\chi(0)=\sigma_\infty=1$. According to  \eq{e8} this solution is stable if $4\sigma_e\sigma_m\ge 1$ and unstable otherwise.
 The stability condition is never satisfied in a medium without magnetic monopoles  $\sigma_m=0$. In this case a small perturbation inevitably drives the chiral conductivity to the only stable stationary solution $\sigma_\infty=0$ resulting in transfer of all helicity into the magnetic field (with monotonically increasing magnetic helicity) and vanishing of $\sigma_\chi$ \cite{Boyarsky:2011uy,Hirono:2015rla,Tuchin:2017vwb}. Essentially, the stability of the $\sigma_\chi(t)=1$ solution  reflects the stability of the chiral medium. 

\medskip 

Thus far in this section we discussed instability of the magnetic helicity. Now we would like to investigate the condition for the magnetic field growth, which is referred to in the literature as the magnetic field instability \cite{Tuchin:2014iua,Manuel:2015zpa,Joyce:1997uy,Boyarsky:2011uy,Kharzeev:2013ffa,Khaidukov:2013sja,Kirilin:2013fqa,Avdoshkin:2014gpa,Akamatsu:2013pjd,Dvornikov:2014uza,Buividovich:2015jfa,Sigl:2015xva,Xia:2016any,Kirilin:2017tdh}. From \eq{e1} and \eq{e4} we derive that 
\ball{e15}
\left(\frac{B}{B_0}\right)^2= \frac{\mathcal{H}_\text{em}}{\mathcal{H}_\text{tot}(0)}\frac{\sigma_\chi}{f(1-f)}\,.
\gal
Taking the time derivative and requiring it to be positive we find 
\ball{e17}
-\frac{d}{d\sigma_\chi}\left[ F(\sigma_\chi)\sigma_\chi\right]+f+F(1-f)<0\,.
\gal
Using \eq{d23} this yields the instability condition 
\ball{e19}
\sigma_\chi-\sigma_m\ln \left[\sigma_\chi +\sqrt{\sigma_\chi^2+(\sigma_e-\sigma_m)^2}\right]-\frac{f}{2}-\frac{1}{2}F(1-f)>0\,.
\gal
This is different from the condition of the magnetic helicity instability $\sigma_\chi^2 > 4 \sigma_e \sigma_m$ derived above (even though \fig{fig:solutions} might hint otherwise). In the limit $\sigma_m\to 0$, Eq.~\eq{e19} reduces to the condition $\sigma_\chi >1/2$ derived in  \cite{Tuchin:2017vwb}.

To conclude this section, we  verify that condition \eq{a23} of the general analysis is satisfied in the FGS model.  This is achieved by identifying 
\ball{e9}
\begin{split}
2\mu =-\lim_{t\to \infty}\dot \gamma(t)=\,& \sigma_e+\sigma_m - \sqrt{(\sigma_e-\sigma_m)^2+\sigma_\infty^2} 
= \frac{4\sigma_e\sigma_m - \sigma^2_\infty}{\sigma_e+\sigma_m + \sqrt{(\sigma_e-\sigma_m)^2+\sigma_\infty^2}}\,.
\end{split}
\gal
It follows that $4(\sigma_e-\mu)(\sigma_m-\mu)= \sigma_\infty^2$ implying that the FGS model has the smallest possible value of $\sigma_\infty$ consistent with \eq{a23}. Also, since,  $\sigma_\infty\le2\sqrt{\sigma_e\sigma_m}$ as we showed beneath \eq{e6}, $\mu$ is positive as required.


\section{Summary and discussion}\label{sec:s}
 
 In this paper we considered classical electrodynamics with Ohm's electric  $\b j_e=\sigma_e \b E$,  magnetic $\b j_m=\sigma_m\b B$ and the anomalous  $\b j_a=\sigma_\chi \b B$ currents. Addition of the magnetic current carried by magnetic monopoles is an extension of the Maxwell-Chern-Simons theory. Similarly to the anomalous current, the magnetic current does not dissipate energy because $\sigma_m$ is invariant under the time-reversal. However, unlike the anomalous current it dissipates the magnetic chirality which can be seen in the chiral anomaly equation \eq{a17}. We argued that in the presence of the magnetic current, the chiral medium exhibits superconductivity when $\sigma_\chi^2< 4\sigma_e\sigma_m$. The macroscopic manifestation of superconductivity is the Meissner effect for magnetic and electric fields,  which indicates the dyonic nature of the condensate. The corresponding London penetration depth is given by \eq{m30}.   
 
We employed the adiabatic approximation to consider the dynamical evolution of the chiral conductivity and magnetic helicity. Our main goal was to understand the properties of the system at the end of the chiral evolution. We found  that the presence of the magnetic current  influences the inverse cascade in a critical way. Whereas at $\sigma_m=0$ the inverse cascade transfers all helicity from the medium to the magnetic field, at finite $\sigma_m$ only a fraction of the medium's helicity can be transferred to the field, while another fraction leaks out. Thus the inverse cascade stops at a finite value of $\sigma_\chi\to \sigma_\infty$ and vanishing magnetic helicity. We analyzed the properties of the stationary states $\sigma_\infty$ and argued that they satisfy the superconductivity condition $\sigma_\chi^2< 4\sigma_e\sigma_m$. We showed that a medium with such chiral conductivity is stable, meaning that a small fluctuation of magnetic helicity decays exponentially with time.  

It is remarkable that the stability condition is identical to the condition for the superconductivity. We conclude that at the end of the inverse cascade the medium reaches a superconducting state that expels electric and magnetic fields. Both electric charges and magnetic monopoles are strongly correlated in such a state.  This observation may explain the strong coupling  of the quark-gluon plasma observed in relativistic heavy-ion collisions (first proposed in \cite{Chernodub:2009rt}),  the most dramatic manifestation of which is its near perfect fluidity \cite{Schafer:2009dj}.

Throughout the paper we treated $\sigma_e$ and $\sigma_m$ as independent quantities. However, they are related to each other through the Dirac quantization condition $eg=N/2$, where $e$,$g$ are electric and magnetic charges and $N$ is an integer.  To establish the precise relationship between $\sigma_e$ and $\sigma_m$ one need to know how these quantities depend on $e$ and $g$. Assuming that the magnetic monopoles are heavier than the electric charges, such dependence is not trivial, but can be computed using the kinetic theory \cite{Arnold:2000dr}.

\acknowledgments
We are grateful to Rebecca Flint, Pieter Maris, Thomas Koschny and James Vary for informative discussions. 
This work  was supported in part by the U.S. Department of Energy under Grant No.\ DE-FG02-87ER40371.



\end{document}